\documentclass[11pt,twoside]{amsart}

\usepackage[hmargin={3cm,3cm},vmargin={3cm,3cm},includehead]{geometry}
\usepackage{cite}
\usepackage{amsmath,amssymb}
\usepackage{amsfonts}
\usepackage{mathrsfs}

\newcommand{\tensprod}{\mathop{\textrm{\Large $\otimes$}}}

\newcommand{\bosn}{\textrm{\scriptsize $\boldsymbol{N}$}}
\newcommand{\fern}{\textrm{\scriptsize $\boldsymbol{M}$}}

\newcommand{\bos}{\boldsymbol{b}}
\newcommand{\fer}{\boldsymbol{f}}

\newcommand{\rop}{\mathsf{R}}

\begin{document}

\title[Hidden structure of $\mathscr{U}_q(D_n^{(1)})$]
{Tetrahedron equations, boundary states and hidden structure of $\mathscr{U}_q(D_n^{(1)})$}%

\author{Sergey M. Sergeev}

\address{Faculty of Informational Sciences and Engineering,
University of Canberra, Bruce ACT 2601}

\email{sergey.sergeev@canberra.edu.au}


\begin{abstract}
Simple periodic $3d\to 2d$ compactification of the tetrahedron
equations gives the Yang-Baxter equations for various evaluation
representations of $\mathscr{U}_q(\widehat{sl}_n)$. In this paper
we construct an example of fixed non-periodic $3d$ boundary
conditions producing a set of Yang-Baxter equations for
$\mathscr{U}_q(D_n^{(1)})$. These boundary conditions resemble a
fusion in hidden direction.
\end{abstract}

\maketitle

The tetrahedron equation can be viewed as a local condition
providing existence of an infinite series of Yang-Baxter
equations. In the applications to quantum groups the method of
tetrahedron equation is a powerful tool for generation of
$R$-matrices and $L$-operators for various ``higher spin''
evaluation representations. This has been demonstrated in
\cite{BS05} for $\mathscr{U}_q(\widehat{sl}_n)$ and in
\cite{supertetrahedron} for super-algebras
$\mathscr{U}_q(\widehat{gl}_{n|m})$.

The main principle producing the cyclic $\widehat{sl}_n$ structure
is the trace in hidden ``third'' direction.  In this paper we
introduce another boundary condition, a certain boundary states
still providing the existence of effective Yang-Baxter equation
and integrability.

We shall start with a short remained of a (super-)tetrahedron
equation and $\widehat{sl}_n$ compactification in their elementary
form. The simplest known tetrahedron equation in the tensor
product of six spaces $B_1\otimes F_2\otimes \cdots \otimes
F_5\otimes B_6$ is
\begin{equation}\label{te}
\rop_{B_1F_2F_3}\rop_{B_1F_4F_5}\rop_{F_2F_4B_6}\rop_{F_3F_5B_6}\;=\;
\rop_{F_3F_5B_6}\rop_{F_2F_4B_6}\rop_{B_1F_4F_5}\rop_{B_1F_2F_3}\;,
\end{equation}
where $F_i=\{|0\rangle,|1\rangle\}_{i}$ is a representation space
of Fermi oscillator
\begin{equation}
\fer^+|0\rangle\;=\;|1\rangle\;,\quad \fer^{-}|1\rangle =
|0\rangle\;.
\end{equation}
Odd operators $\fer_i^{\pm}$ in different components $i$ of their
tensor product anti-commute and $(\fer_i^{\pm})^2=0$. It is
convenient to introduce projectors
\begin{equation}\label{B-basis}
\fern_i^{}=\fer_i^+\fer_i^-\;,\quad
\fern_i^{0}=\fer_i^-\fer_i^+\;,\quad
[\fer_i^+,\fer_i^-]_+=\fern_i^{0}+\fern_i^{}=1\;.
\end{equation}
Operator $\fern_i^{0}$ is the projector to vacuum, $\fern_i^{}$ is
the occupation number and $\fern^0\fern=0$.

Space $B_i$ stands for representation space of $i$-th copy of
$q$-oscillator,
\begin{equation}
\bos^+\bos^-=1-q^{2\bosn}\;,\quad
\bos^-\bos^+=1-q^{2\bosn+2}\;,\quad q^{\bosn}\bos^{\pm}=\bos^{\pm}
q^{\bosn\pm 1}\;.
\end{equation}
In this paper we imply the unitary Fock space representation,
$(\bos^-)^\dagger=\bos^+$, defined by
\begin{equation}
\bosn|n\rangle = |n\rangle n\;,\quad \bos^-|0\rangle = 0\;,\quad
|n\rangle \;=\; \frac{\bos^{+n}}{\sqrt{(q^2;q^2)_n}}\,
|0\rangle\;,\quad n\geq 0\;,
\end{equation}
where $(x;q^2)_n=(1-x)(1-q^2x)\cdots (1-q^{2n-2}x)$. In terms of
creation, annihilation and occupation number operators the
$\rop$-matrices in (\ref{te}) are given \cite{supertetrahedron} by
\begin{equation}\label{raff}
\rop_{B_1F_2F_3}=\fern_2^{0}\fern_3^{0} -
q^{\bosn_1+1}\fern_2^{}\fern_3^{0} +
q^{\bosn_1}\fern_2^{0}\fern_3^{}-\fern_2^{}\fern_3^{} +
\bos_1^-\fer_2^+\fer_3^- - \bos_1^+\fer_2^-\fer_3^+
\end{equation}
and
\begin{equation}\label{rffa}
\rop_{F_1F_2B_3}=\fern_1^{0}\fern_2^{0} +
\fern_1^{}\fern_2^{0}q^{\bosn_1+1} -
\fern_1^{0}\fern_2^{}q^{\bosn_1}-\fern_2^{}\fern_3^{} +
\fer_1^+\fer_2^-\bos_3^- - \fer_1^-\fer_2^+\bos_3^+\;.
\end{equation}
Both operators $\rop$ are unitary roots of unity. The constant
tetrahedron equation (\ref{te}) can be verified in the operator
language straightforwardly.

Define next the ``monodromy'' of $\rop$-matrices as the ordered
product
\begin{equation}
\rop_{\Delta_n(B_1F_2),F_3}\;=\;
\rop_{B_{1:1}F_{2:1}F_3}\rop_{B_{1:2}F_{2:2}F_3}\cdots
\rop_{B_{1:n}F_{2:n}F_3}\;\leftrightharpoons\;\prod_{j=1..n}^\curvearrowright
\rop_{B_{1:j}F_{2:j}F_3}\;.
\end{equation}
Here the convenient ``co-product'' notation stands for a tensor
power of corresponding spaces,
\begin{equation}
\Delta_n(B_1)\;=\;\tensprod_{j=1}^n B_{1:j}\;,\quad
\Delta_n(F_2)\;=\;\tensprod_{j=1}^n F_{2:j}\;.
\end{equation}
The repeated use of (\ref{te}) provides
\begin{equation}\label{te2}
\begin{array}{l}
\rop_{\Delta_n(B_1F_2),F_3}\rop_{\Delta_n(B_1F_4),F_5}\rop_{\Delta_n(F_2F_4),B_6}\rop_{F_3F_5B_6}\\
[1mm]
\phantom{xxxxxxxxxxxxxxxxxxxxx}\;=\;
\rop_{F_3F_5B_6}\rop_{\Delta_n(F_2F_4),B_6}\rop_{\Delta_n(B_1F_4),F_5}\rop_{\Delta_n(B_1F_2),F_3}\;.
\end{array}
\end{equation}
Note the conservation laws:
\begin{equation}
v^{-\fern_3}u^{-\fern_5}\left(\frac{u}{v}\right)^{\bosn_6}\rop_{F_3F_5B_6}\;=\;
\rop_{F_3F_5B_6}
v^{-\fern_3}u^{-\fern_5}\left(\frac{u}{v}\right)^{\bosn_6}\;.
\end{equation}
Multiplying (\ref{te2}) by the $u,v$-term in $F_3\otimes
F_5\otimes B_6$ and by $\rop_{F_3F_5B_6}^{-1}$,  and making then
the traces over $F_3\otimes F_5\otimes B_6$, we come to the
Yang-Baxter equation\
\begin{equation}\label{ybe}
L_{\Delta_n(B_1F_2)}(v)L_{\Delta_n(B_1F_4)}(u)
R_{\Delta_n(F_2F_4)}(u/v)\;=\; R_{\Delta_n(F_2F_4)}(u/v)
L_{\Delta_n(B_1F_4)}(u)L_{\Delta_n(B_1F_2)}(v)\;,
\end{equation}
where
\begin{equation}\label{2d-R}
L_{\Delta_n(B_1F_2)}(v)=\mathop{\mathrm{Str}}_{F_3} \left(
v^{-\fern_3} \rop_{\Delta_n(B_1F_2),F_3}\right),\quad
R_{\Delta_n(F_2F_4)}(w) = \mathop{\mathrm{Tr}}_{B_6}\left(
w^{\bosn_6} \rop_{\Delta_n(F_2F_4),B_6}\right).
\end{equation}
This is the case of $\mathscr{U}_q(\widehat{sl}_n)$.
Two-dimensional $R$-matrices (\ref{2d-R}) have the centers
\begin{equation}
J_i=\sum_{j=1}^n \fern_{i:j}\quad \textrm{for fermions and }\;
J_1=\sum_{j=1}^n \bosn_{1:j}\quad \textrm{for bosons.}
\end{equation}
Irreducible components of $R$-matrices and $L$-operators
(\ref{2d-R}) correspond to fixed values of $J_i$. In particular,
$\Delta_n(F)$ is the sum of all antisymmetric tensor
representations of $sl_n$,
\begin{equation}
\dim\Delta_n(F)\;=\;2^n\;=\;\sum_{k=0}^n \frac{n!}{k!(n-k)!}\;.
\end{equation}

\bigskip

The Dirac spinor representation of $D_n$ has the same dimension
$2^n$, it is the direct sum of two irreducible Weyl spinors with
dimensions $2^{n-1}$. It is evident intuitively, the structure of
$D_n$ will appear if the total occupation number $J$ of
$\Delta_n(F)$ is not a center of $L$-operators and $R$-matrices,
but all operators preserve just the parity of $J$. Also, since the
dimension of vector representation of $D_n$ is $2n$, we need to
double the number of bosons.

Consider now two copies of (\ref{te}) and further of (\ref{te2})
glued in the ``second'' direction. This consideration keeps the
desired space $\Delta_n(F)$ and doubles the number of bosons. The
repeated use of (\ref{te}) provides
\begin{equation}\label{te3}
\begin{array}{l}
\rop_{\Delta(B_1)F_2\Delta(F_3)} \rop_{\Delta(B_1)F_4\Delta(F_5)}
\rop_{F_2F_3B_6} \rop_{\Delta'(F_3F_5)B_6} \\
[1mm]
\phantom{xxxxxxxxxxxxxxxxx} = \rop_{\Delta'(F_3F_5)B_6}
\rop_{F_2F_3B_6} \rop_{\Delta(B_1)F_4\Delta(F_5)}
\rop_{\Delta(B_1)F_2\Delta(F_3)},
\end{array}
\end{equation}
where
\begin{equation}\label{double-r}
\rop_{\Delta(B_1)F_2\Delta(F_3)}\;=\;
\rop_{B_1^{}F_2^{}F_3^{}}\rop_{B_1'F_2^{}F_3'}\;\;\;\textrm{and}\;\;\;
\rop_{\Delta'(F_3F_5)B_6}=\rop_{F_3'F_5'B_6^{}}\rop_{F_3^{}F_5^{}B_6^{}}\;.
\end{equation}
The key observation is the existence of a family of eigenvectors
of operator $\rop_{\Delta'(F_3F_5)B_6}$:
\begin{equation}\label{psivectros}
\rop_{\Delta'(F_3F_5)B_6}|\psi_{\Delta(F_3)}(v)\psi_{\Delta(F_5)}(u)\psi_{B_6}(u/v)\rangle\;=\;
|\psi_{\Delta(F_3)}(v)\psi_{\Delta(F_5)}(u)\psi_{B_6}(u/v)\rangle\;,
\end{equation}
where
\begin{equation}
\Delta(F)=F'\otimes F\;,\quad
|\psi_{\Delta(F)}(v)\rangle\;=\;(1+v^{-1}\fer^{+\prime}\fer^{+})|0\rangle\;,
\end{equation}
and in the unitary basis (\ref{B-basis})
\begin{equation}
\langle 2k+1|\psi_B(w)\rangle=0\;,\quad \langle
2k|\psi_B(w)\rangle= w^{k}
\sqrt{\frac{(q^{4k+4};q^4)_\infty}{(q^{4k+2};q^4)_\infty}}\;.
\end{equation}
The normalization of $\psi_B$ is given by
\begin{equation}\label{normalization}
\langle \overline{\psi}_B(w)|(\bos^{\pm})^{2m}|\psi_B(w)\rangle
\;=\; w^m \frac{(q^{2+4m}w^2;q^4)_\infty}{(w^2;q^4)_\infty}\;.
\end{equation}
Considering now a length-$n$ chain of (\ref{te3}) in the ``third''
direction and applying vectors $\psi_{\Delta(F_3)}(u)$,
$\psi_{\Delta(F_5)}(v)$ and $\psi_B(u/v)$, we come to the
Yang-Baxter equation
\begin{equation}
\begin{array}{l}
L_{\Delta_n(\Delta(B_1)F_2)}(v) L_{\Delta_n(\Delta(B_1)F_4)}(u)
R_{\Delta_n(F_2F_4)}(u/v) \\
[1mm]
\phantom{xxxxxxxxxxxxx} = R_{\Delta_n(F_2F_4)}(u/v)
L_{\Delta_n(\Delta(B_1)F_4)}(u) L_{\Delta_n(\Delta(B_1)F_2)}(v)
\end{array}
\end{equation}
without trace construction:
\begin{equation}\label{Lop-D}
L_{\Delta_n(\Delta(B_1)F_2)}(v)\;=\;\langle\overline{\psi}_{\Delta(F_3)}(v)|
\rop_{\Delta_n(\Delta(B_1)F_2),\Delta(F_3)}|\psi_{\Delta(F_3)}(v)\rangle
\end{equation}
and
\begin{equation}\label{Rop-D}
R_{\Delta_n(F_2F_4)}(w)\;=\; \langle\overline{\psi}_{B_6}(w)|
\rop_{\Delta_n(F_2F_4),B_6} |\psi_{B_6}(w)\rangle\;.
\end{equation}
Matrix elements of $R_{\Delta_n(F_2F_4)}(w)$ can be calculated
with the help of (\ref{normalization}) and similar identities. The
invariants of $L$-operator (\ref{Lop-D}) and $R$-matrix
(\ref{Rop-D}) are: the parity of $J_2=\sum\fern_{2:j}$, similar
parity of $J_4$ and
\begin{equation}
J_1\;=\;\sum_{j=1}^n (\bosn_{1:j}^{}-\bosn_{1:j}')\;.
\end{equation}
A choice of different spectral parameters in bra- and ket-vectors
in (\ref{Lop-D},\ref{Rop-D}) is equivalent to the choice of equal
spectral parameters by means of a gauge transformation.

The structure of $D_n$ representation ring can be verified
explicitly by a direct calculation of matrix elements of
$R$-matrix (\ref{Rop-D}) for small $n$ and check of factor powers
of $\det(\lambda-R)$.

As to $2n$-bosons space, irreducible components of
$\Delta_n(\Delta(B_1))$ are in general infinite dimensional.
However, a choice of Fock and anti-Fock space representations,
$\textrm{Spectrum}(\bosn_{1:j})=0,1,2,\dots$ and
$\textrm{Spectrum}(\bosn_{1:j}')=-1,-2,-3,\dots$, makes
$\Delta_n(\Delta(B_1))$ a direct sum of symmetric tensors of
$O(2n)$.

\bigskip

The main result of this paper is a step forward to a
classification of \emph{integrable boundary conditions} in
three-dimensional models. At least two scenarios are hitherto
known: quasi-periodic boundary condition (\ref{2d-R}) and the
boundary states condition (\ref{Lop-D},\ref{Rop-D}). These
conditions can be imposed for a layer-to-layer transfer matrix in
different directions independently. In both scenarios the spectral
parameters of effective two-dimensional models reside the
boundary. Also, the boundary admits twists making the quantum
groups classification inapplicable \cite{BoseGas}. It worth noting
one more possible scenario of integrable boundary conditions: yet
unknown $3d$ reflection operators satisfying the tetrahedron
reflection equations \cite{TRE}.

\noindent\textbf{Acknowledgements.} I would like to thank all
staff of the Faculty of Information Science for their support.


\begin{thebibliography}{1}

\bibitem{BS05}
V. V. Bazhanov and S. M. Sergeev, \emph{Zamolodchikov's
tetrahedron equation and hidden structure of quantum groups}, J.
Phys. A \textbf{39} (2006), no.~13, 3295--3310

\bibitem{TRE}
A.~P. Isaev and P.~P. Kulish, \emph{Tetrahedron reflection
equations}, Modern Phys. Lett. A \textbf{12} (1997), no.~6,
427--437

\bibitem{BoseGas}
S.~Sergeev, \emph{Ansatz of {H}ans {B}ethe for a two-dimensional
lattice {B}ose gas}, J. Phys. A \textbf{39} (2006), no.~12,
3035--3045

\bibitem{supertetrahedron}
S.~M. Sergeev, \emph{{S}uper-tetrahdera and super-algebras},
arXiv:0805.4653, 2008.

\end{thebibliography}

\end{document}